# A Triple-Band Bandpass Filter Using Square Open-Loop Resonators

Rashmi Ravi
*School of Engineering*
*University of Greenwich*
*Chatham Maritime, Kent, UK*
rr8438i@gre.ac.uk

Bhaskarareddy N C
*School of Engineering*
*University of Greenwich*
*Chatham Maritime, Kent, UK*
nb6621j@gre.ac.uk

Augustine O. Nwajana
*School of Engineering*
*University of Greenwich*
*Chatham Maritime, Kent, UK*
A.O.Nwajana@greenwich.ac.uk

***Abstract—****The rapid advancement of multi-band wireless communication systems has driven demand for compact, high-performance multi-band bandpass filters (BPFs) capable of isolating specific frequency bands within a wider spectrum. This paper reviews the current state-of-the-art in multi-band filter design and implementation techniques and presents the design, simulation, and characterisation of a prototype triple-band bandpass filter to validate one of the investigated techniques. A triple-band BPF operating at 2.1 GHz, 2.2 GHz, and 2.3 GHz is designed using Keysight ADS software and implemented on Rogers RT/Duroid 6010LM substrate (dielectric constant = 10.7, loss tangent = 0.0023, thickness = 1.27 mm). The design employs nine square open-loop resonators three per passband with a characteristic impedance of 50 Ω and a fractional bandwidth of 6%. Simulation results demonstrate insertion losses of 0.674 dB, 0.976 dB, and 1.314 dB, and return losses of 16.785 dB, 33.609 dB, and 17.162 dB across the three passbands respectively. The 2.2 GHz centre frequency is achieved with high precision, confirming the effectiveness of the square open-loop resonator transformation technique for compact multi-band filter design in modern wireless communication systems.*

## I. INTRODUCTION

In the rapidly evolving landscape of wireless communication technology, the demand for smaller and more efficient devices has become increasingly prevalent. Modern communication systems such as 4G LTE, UMTS, 5G, Wi-Fi, and Bluetooth are required to operate simultaneously across multiple frequency bands, necessitating components that can handle multiple channels within a single compact device [17]. Bandpass filters (BPFs) are fundamental passive components in radio frequency (RF) and microwave systems, responsible for selectively transmitting signals within specified frequency bands while rejecting unwanted signals outside those bands. Figure (1) shows the conventional approach to achieving multi-band operation involves cascading multiple single-band filters, which not only increases the physical size of the system but also introduces higher insertion losses and greater complexity in the RF front-end [5].

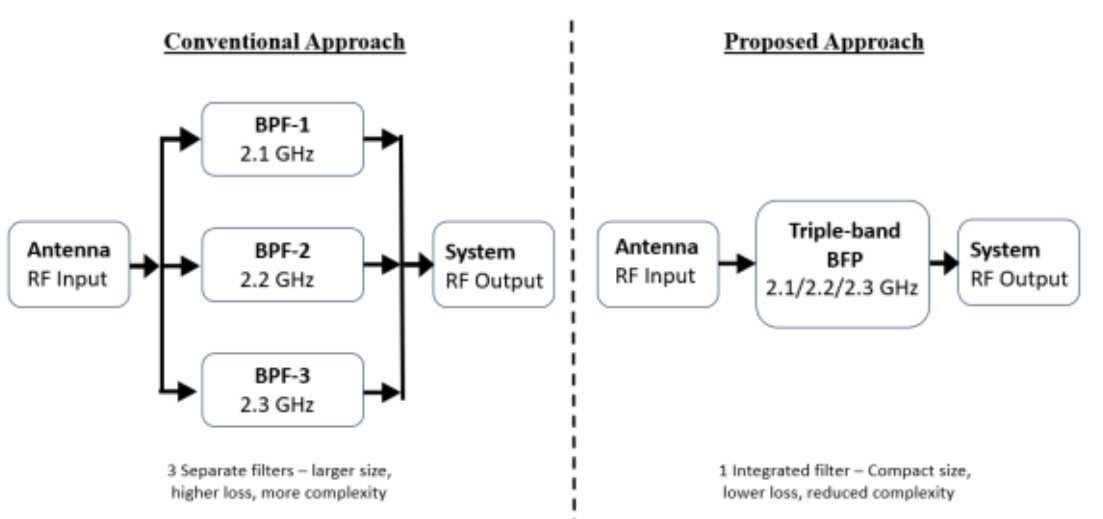

*Fig. 1. Conventional cascaded single-band filters versus the proposed integrated triple-band filter.*

There is a growing interest in the design of multi-band filters as a single integrated component in communication systems. Several authors have proposed various techniques for implementing this type of filter. Dual-mode resonator-based approaches have been widely explored due to their ability to reduce the physical size of filter devices while enhancing overall performance [15], [20]. Substrate integrated waveguide (SIW) technology has also been deployed in multi-band filter design, offering compatibility with standard PCB manufacturing processes [4], [16]. However, SIW-based devices are relatively bulky when compared to microstrip implementations and hence occupy a larger footprint. Stepped impedance resonators (SIR) have been employed to achieve multi-band responses with spurious band rejection, though results reported in the literature suggest that the intended performance is not always fully attained [9], [10]. Hairpin resonators and open-loop ring resonators have also been extensively used in multi-band filter implementations, offering compact size and relatively simple design procedures [11], [18].

The triple-band BPF proposed in this paper is achieved through the transformation of a single-band BPF and implemented using microstrip square open-loop resonator technology for compactness and ease of integration with other devices in any communication system. The proposed design operates at centre frequencies of 2.1 GHz, 2.2 GHz, and 2.3 GHz, employing nine square open-loop resonators three resonators per passband on Rogers RT/Duroid 6010LM substrate with a dielectric constant of 10.7, a loss tangent of 0.0023, and a thickness of 1.27 mm. The design is first validated through circuit modelling using a capacitor-only network before being extended to the full microstrip layout and simulated using Keysight Advanced Design System (ADS) electromagnetic simulation software.

## II. FILTER DESIGN AND IMPLEMENTATION

### A. Theoretical Circuit Configuration

The procedure for achieving the triple-band bandpass filter begins with the transformation of the standard normalised prototype Chebyshev lowpass filter element values of g0 =

g4 = 1, g1 = g3 = 0.8516, and g2 = 1.1032 to realise a single-band bandpass filter at the desired centre frequency of 2.2 GHz, fractional bandwidth (FBW) of 6%, and input/output characteristic impedance (Z0) of 50 Ω. The single-band BPF transformation is based on the principles reported in [7], using a capacitor-only circuit model. The final circuit configuration of the single-band BPF is first established before the triple-band configuration is derived from it.

The circuit component values for the single-band BPF at 2.2 GHz are determined using the following design formulations [5], [6]:

$$L_1 = \frac{Z*FBW}{g_1*2\pi f} \;,\; C_1 = \frac{g_1}{Z*2\pi f*FBW} \tag{1}$$

$$J_{01} = J_{34} = \sqrt{\frac{g_0}{Z}} \;,\; J_{12} = J_{23} = \sqrt{\frac{g_1 g_2}{Z}} \tag{2}$$

$$C_{m,01} = C_{m,34} = \frac{J_{01}}{2\pi f} \;,\; C_{m,12} = C_{m,23} = \frac{J_{12}}{2\pi f} \tag{3}$$

The single-band bandpass filter at 2.2 GHz was modelled using a capacitor-only circuit, yielding a series inductance $L_1$ of 0.2549 nH and capacitance $C_1$ of 20.540 pF. The port coupling capacitances $C_{m,01} = C_{m,34}$ were calculated as 1.447 pF, while the inter-resonator coupling capacitances $C_{m,12} = C_{m,23}$ were found to be 1.266 pF.

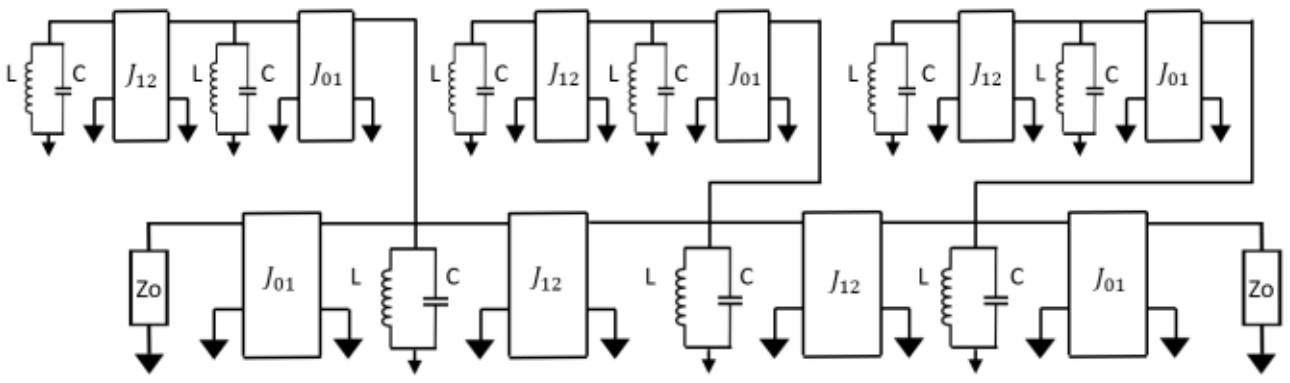

*Fig. 3. Schematic circuit configuration of the proposed triple-band BPF.*

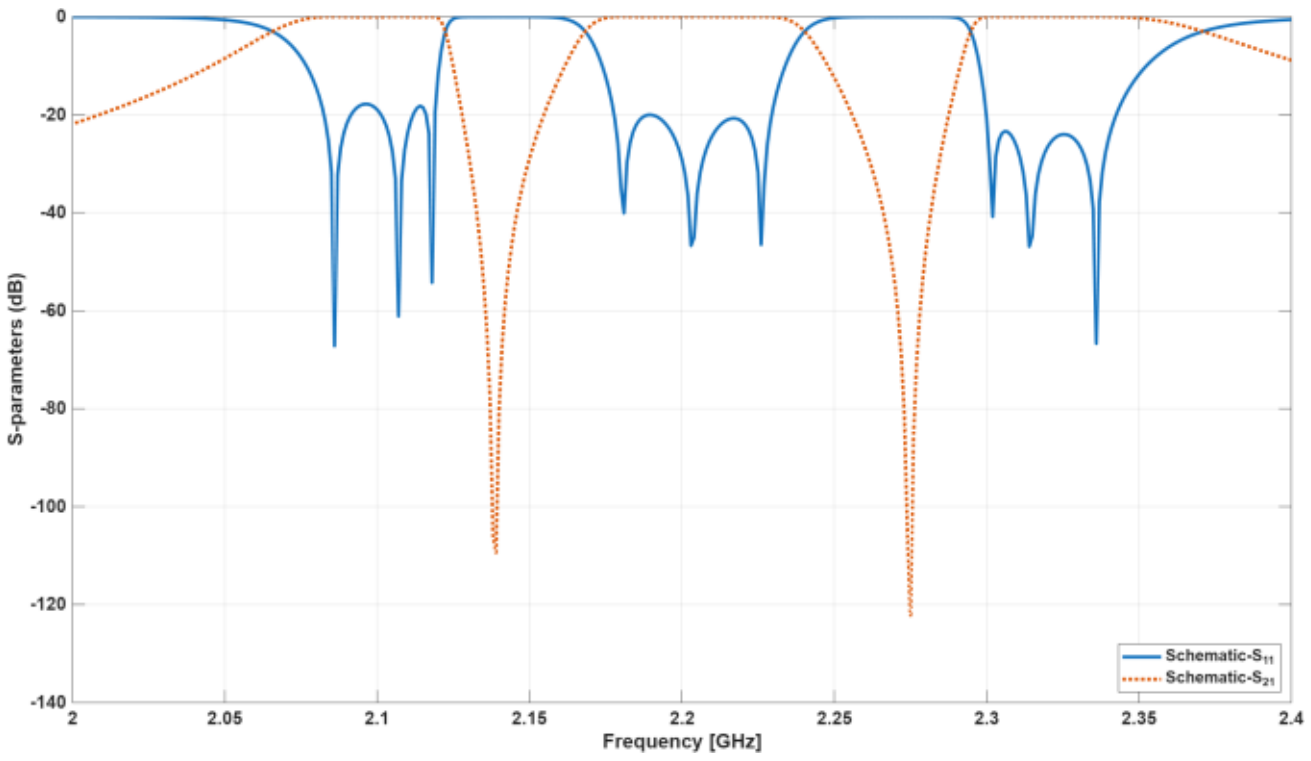

*Fig. 4. Schematic simulation response of the proposed triple-band BPF.*

The schematic simulation of the proposed triple-band BPF yields three passbands centred at 2.1, 2.2, and 2.3 GHz with fractional bandwidths (FBWs) of 5.6%, 7.3%, and 7.6%, respectively. Notably, two transmission zeros are observed at 2.139 GHz and 2.275 GHz, located in the stopband regions between adjacent passbands. These transmission zeros arise from the destructive interference of signal paths through the coupled square open-loop resonators and contribute significantly to the selectivity of the filter by providing sharp attenuation between the three passbands, thereby improving inter-band isolation.

The triple-band BPF is then derived from the single-band BPF by designing two additional filter branches at 2.1 GHz and 2.3 GHz, following the same transformation procedure applied to the 2.2 GHz single-band filter. The three individual single-band filters, centred at 2.1 GHz, 2.2 GHz, and 2.3 GHz respectively, are then coupled together using microstrip coupling to form the complete nine-pole triple-band BPF. This approach is consistent with the single-to-multi-band transformation technique reported in [1], where a single-band BPF is extended to produce additional passbands by retaining a similar mutual coupling network between each pair of resonators across the additional filter branches.

The theoretical mutual coupling coefficient M and external quality factor Qe between resonators are determined using Equations (4) and (5) respectively, as reported in [7]:

$$M = \frac{FBW}{\sqrt{(g_1 g_2)}} \tag{4}$$

$$Q_e = \frac{g_0 g_1}{FBW} \tag{5}$$

The theoretical mutual coupling coefficient K was calculated as 0.0619, and the external quality factor Qe was determined to be 14.1933. These values were derived from the Chebyshev prototype element values and the specified fractional bandwidth of 6% and serve as the target parameters for the subsequent practical extraction of coupling distances and tapping lengths in the microstrip layout.

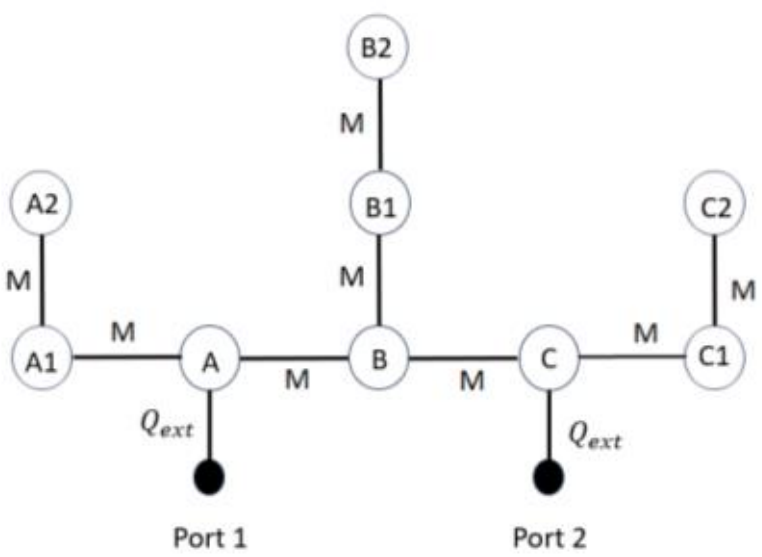

*Fig. 5. Coupling arrangement for the proposed triple-band BPF showing the three coupled filter branches.*

### *B. Microstrip Layout and Practical Implementation*

The square open-loop resonator structure is employed in the implementation of the proposed triple-band bandpass filter due to its simplicity, compact size, and ease of integration with other microwave components, when compared to many other popular microstrip resonator structures. The width and guided wavelength for designing the square open-loop resonator are determined following the procedure reported in [3]. ADS electromagnetic (EM) simulation software is used to construct and simulate the layout of the triple-band filter model. The layout is created on Rogers RT/Duroid 6010LM substrate with a dielectric constant of 10.7, a thickness of 1.27 mm, and a loss tangent of 0.0023. The square open-loop resonator is designed to resonate at the filter centre frequency of 2.2 GHz. The microstrip line width is determined as w = 1.14 mm for the specified substrate parameters and characteristic impedance of 50 Ω.

The guided wavelength at 2.2 GHz on the RT/Duroid 6010LM substrate was determined to be 51.077 mm, giving a half-wavelength of 25.5385 mm and an eighth-wavelength resonator side length of 6.3846 mm. The microstrip line width was calculated as w = 1.14 mm to satisfy the 50 Ω characteristic impedance condition for the given substrate parameters.

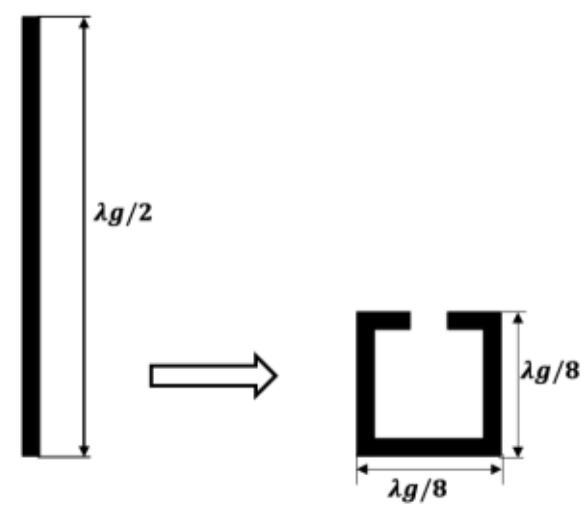

*Fig. 6. Practical dimensions of the proposed square open-loop resonator.*

The proposed triple-band BPF employs nine square open-loop resonators in total, three for each of the three passbands at 2.1, 2.2, and 2.3 GHz respectively. Microstrip coupling is used to connect all three filter branches together. The coupling arrangement shown in Fig. 5 forms the design foundation for the complete triple-band filter layout.

As described in the theoretical circuit configuration, it is necessary to determine the practical mutual coupling coefficient and the external quality factor for the proposed filter. The mutual coupling coefficients are extracted based on three different physical orientations of the square open-loop resonator pairs, as the orientation of the resonator openings relative to each other determines the type of coupling achieved. The practical mutual coupling coefficient is extracted in each case by continuously adjusting the separation distance, s, between the two resonators and simulating until the practical coupling coefficient value, calculated using Equation (6), matches the theoretical value of K = 0.0619.

$$k = \frac{(f_2^2 - f_1^2)}{(f_2^2 + f_1^2)} \qquad (6)$$

When one resonator is placed above the other, the extracted coupling spacing values from ADS EM simulation at 2.2 GHz are s = 1.06 mm, s = 0.75 mm, and s = 1.255 mm for the three resonator orientations, corresponding to openings facing opposite vertical directions, openings facing each other, and base sides facing each other, respectively.To extract the external quality factor, the tapping distance, t, is continuously adjusted and simulated until the practical external quality factor value, calculated using Equation (7), matches the theoretical value of Qe = 14.1933. The centre frequency f0 and the two frequencies f1 and f2 at the 3 dB points below the transmission peak are identified from the simulation response. The practical tapping distance obtained from the ADS EM simulation at 2.2 GHz is t = 0.84 mm.

$$Q_{\text{ext}} = \frac{f_0}{(f_2 - f_1)} \qquad (7)$$

The practical separation distances between square open-loop resonator pairs were extracted through ADS EM simulation by iterating until the simulated coupling coefficient matched the theoretical value of K = 0.0619. For the orientation with one resonator opening upward and one downward, the separation distance was s = 1.06 mm; for both openings directly facing each other, s = 0.75 mm; and for both base sides facing each other with openings pointing outward, s = 1.255 mm. The tapping distance for matching the theoretical external quality factor of Qe = 14.1933 was determined to be t = 0.84 mm.

The physical dimensions of the complete triple-band filter microstrip layout are shown in Fig. 7. The layout comprises nine square open-loop resonators arranged and coupled together across the three filter branches at 2.1, 2.2, and 2.3 GHz respectively.

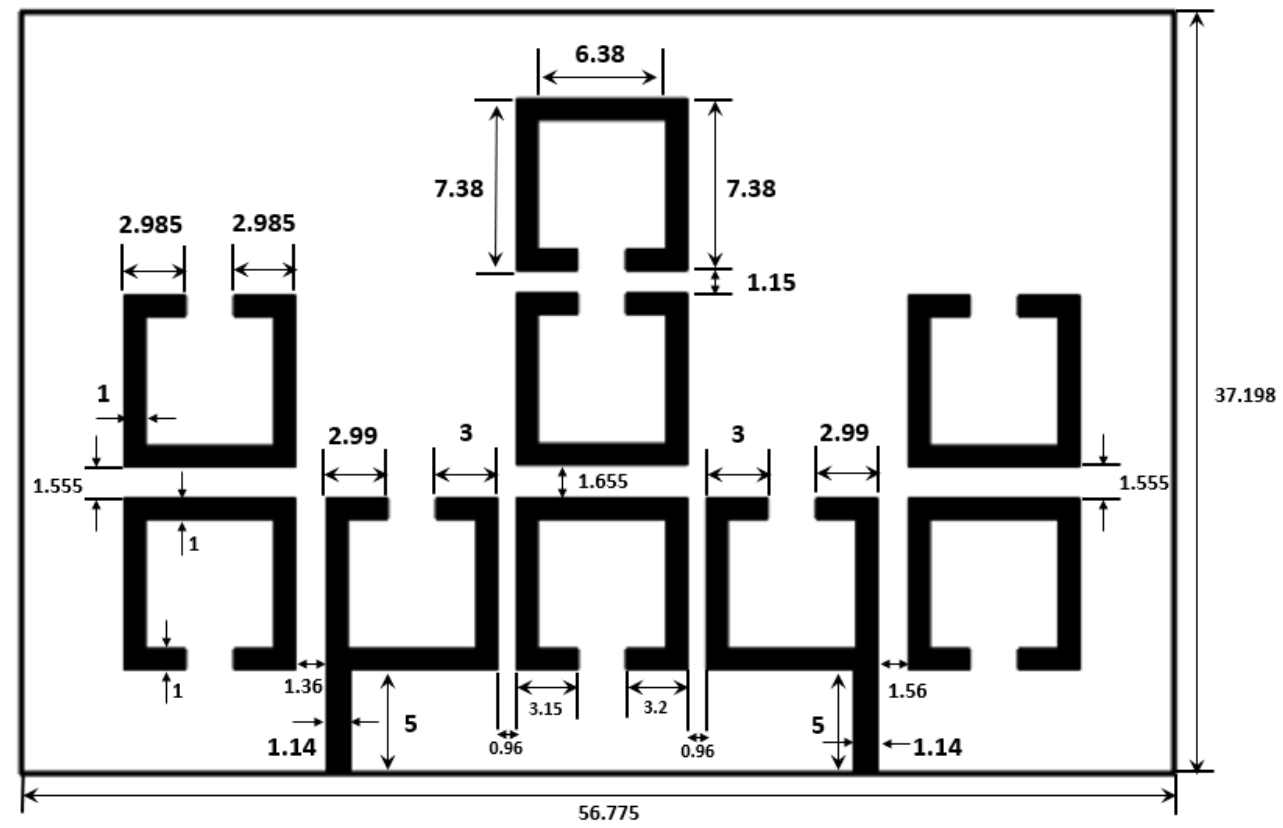

*Fig. 7. Microstrip layout of the proposed triple-band BPF with physical dimensions in mm.*

## III. RESULTS AND DISCUSSION

### A. Results

This section presents and discusses the simulation results for the proposed triple-band BPF. The filter, implemented using microstrip square open-loop resonators, was developed from the transformation of a single-band BPF at 2.2 GHz and then extended to produce the additional passbands at 2.1 GHz and 2.3 GHz. The EM simulation result is presented in Fig. 8.

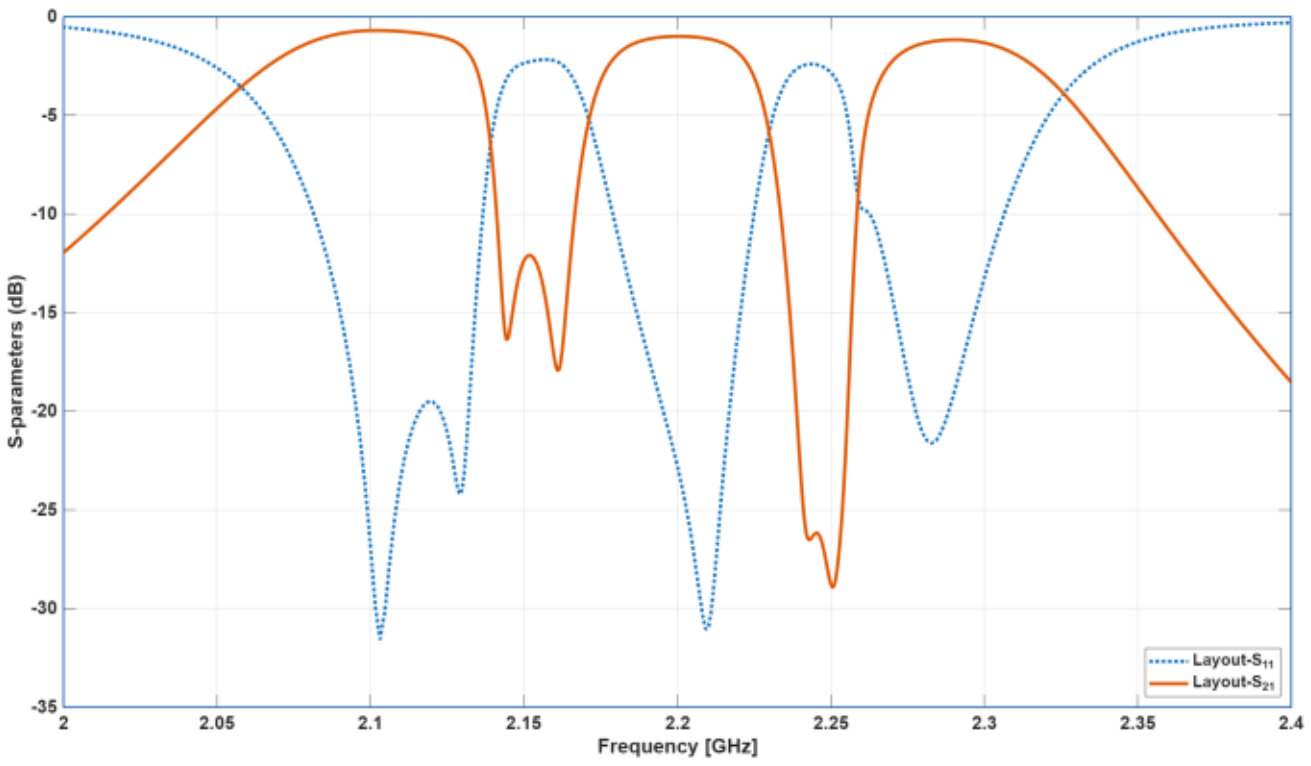

*Fig. 8. EM simulation response of the proposed triple-band BPF showing all three passbands at 2.1, 2.2, and 2.3 GHz.*

The EM simulation response confirms that the proposed filter successfully transmits at all three specified centre frequencies of 2.1, 2.2, and 2.3 GHz simultaneously, validating the nine-resonator microstrip layout and the coupling parameters extracted from the ADS EM simulation procedure.

The schematic and EM simulation responses show good agreement across all three passbands, confirming the accuracy of the design formulations, circuit modelling, and microstrip layout implementation adopted in this research.

The EM simulation results confirm successful triple-band operation at all three target centre frequencies. Insertion losses of 0.674 dB, 0.976 dB, and 1.314 dB were achieved at 2.1, 2.2, and 2.3 GHz respectively, while return losses of 16.785 dB, 33.609 dB, and 17.162 dB were recorded across the three passbands. The 2.2 GHz centre band exhibits the highest return loss, exceeding the standard 20 dB benchmark, owing to the entire design being centred and optimised at that frequency.

### *B. Discussion*

#### *1) Insertion Loss Analysis:*

The proposed triple-band BPF achieves insertion losses of 0.674 dB, 0.976 dB, and 1.314 dB at the first, second, and third passbands respectively. These values are all within acceptable limits for microstrip filter implementations and confirm low signal attenuation across all three passbands. However, a clear progressive increase in insertion loss is observed from the first to the third passband. The primary cause is the conductor and dielectric loss of the substrate; the RT/Duroid 6010LM substrate has a loss tangent of 0.0023, which introduces dielectric loss that increases with frequency from 2.1 GHz to 2.3 GHz. A second factor is the coupling complexity of the nine-resonator arrangement, where each coupling junction introduces additional loss due to radiation, surface-wave excitation, and junction discontinuities that become more pronounced at higher frequency. A third factor is the close proximity of the three passbands within the narrow 200 MHz span, causing partial overlap of inter-resonator coupling fields and slight frequency-pulling effects that are most pronounced at the 2.3 GHz passband.

#### *2) Return Loss Analysis:*

The proposed filter achieves return losses of 16.785 dB, 33.609 dB, and 17.162 dB at the first, second, and third passbands respectively. The outstanding return loss of 33.609 dB at the 2.2 GHz centre band significantly exceeds the reference 20 dB standard and demonstrates exceptional impedance matching. This is directly attributed to the fact that the entire design procedure the Chebyshev prototype element values, the component calculations, the resonator dimensions, and the coupling and external-Q extraction is centred and optimised at 2.2 GHz. The slightly lower return losses at 2.1 GHz and 2.3 GHz arise because these branches are derived from the 2.2 GHz design by scaling the resonator dimensions rather than being independently optimised, introducing small deviations in the coupling coefficient and external quality factor. The slight asymmetry between 2.1 GHz and 2.3 GHz is attributed to the asymmetric frequency-pulling effects of the dominant central 2.2 GHz branch on the two outer branches.

#### *3) Schematic vs EM Simulation Comparison:*

The good agreement between the schematic and EM simulation responses confirms the validity of the circuit-modelling approach. The small deviations observed are attributed to the idealised nature of the circuit model, which does not account for microstrip discontinuities such as open ends, T-junctions, bends, and cross junctions present in the physical layout. These introduce additional parasitic reactances in the EM simulation that are absent in the ideal circuit model, causing small shifts in the centre frequencies and slight differences in the loss values. These deviations are consistent with similar microstrip filter designs reported by [1] and [2] and are considered acceptable for this type of implementation.

### *C. Performance Comparison*

The proposed triple-band BPF is compared to the current state-of-the-art in Table I. The compared works include dual-band designs using asymmetrical parallel-coupled lines [8], stub-loaded structures [9], and stepped impedance resonators [10], as well as suspended coplanar waveguide-microstrip hybrid structures [12], dual-mode dielectric waveguide resonators [13], SIW multi-mode resonators [14], multilayer SIW cavity filters [19], and the single-to-dual-band transformation approach of [1] and the filtering power divider of [2]. The proposed design is the only work among those compared to achieve triple-band operation from a single-band transformation technique, to demonstrate three simultaneous passbands within a narrow 200 MHz span, and to achieve a return loss exceeding 33 dB at the centre frequency. The insertion-loss values are higher than the dual-band designs of [10] and [1], which is expected given the nine-resonator three-band implementation; this trade-off is considered acceptable given the advancement in filter functionality.

TABLE I
PERFORMANCE COMPARISON WITH RELEVANT LITERATURE

| Ref. | f0 (GHz) | Order | IL (dB) | RL (dB) |
|---|---|---|---|---|
| [1] | 1.35/1.45 | 2/2 | 0.1/0.1 | 24/18 |
| [2] | 2.6 | 3 | 4.77 | 15.5 |
| [8] | 2.3/3.7 | 2/2 | 1.1/1.9 | 23/21 |
| [9] | 0.9/2.5 | 2/2 | 0.3/0.9 | 20/12 |
| [10] | 2.9/4.4 | 2/2 | 0.1/0.1 | 17/17 |
| This work | 2.1/2.2/2.3 | 3/3/3 | 0.674/0.976/1.314 | 16.8/33.6/17.2 |

### *D. Novelty and Uniqueness of the Proposed Work*

The proposed triple-band BPF presents several distinctive contributions. First, it demonstrates a straightforward and reproducible transformation technique that derives a nine-pole triple-band BPF directly from a single-band design; while [1] demonstrated a single-to-dual-band transformation using U-shaped resonators, the present work extends this to triple-band operation using square open-loop resonators. Second, the design achieves three poles per band across three simultaneous passbands nine poles in a single integrated device contributing to sharper selectivity and better inter-band rejection than typical dual-band designs. Third, it achieves a return loss of 33.609 dB at the centre frequency, the highest among the compared works. Fourth, the three passbands are positioned within a narrow 200 MHz span (100 MHz separation), which is technically challenging and directly relevant to closely spaced 4G LTE, UMTS, and sub-3 GHz 5G channels. Finally, the entirely microstrip implementation on RT/Duroid 6010LM ensures compatibility with standard PCB manufacturing and a compact footprint, and the complete validated methodology is fully reproducible and extensible to other bands and higher orders.

## IV. CONCLUSION AND FUTURE WORK

A triple-band bandpass filter derived from the transformation of a single-band bandpass filter has been designed, implemented, and simulated using microstrip square open-loop resonator technology. The proposed filter operates at 2.1, 2.2, and 2.3 GHz, employing nine square open-loop resonators three per passband on RT/Duroid 6010LM substrate, and is validated through Keysight ADS EM simulation with good agreement between schematic and EM responses. The filter achieves insertion losses of 0.674, 0.976, and 1.314 dB, and return losses of 16.785, 33.609, and 17.162 dB across the three bands. The 2.2 GHz centre frequency is achieved with high precision, and the 33.609 dB return loss at the centre band exceeds the 20 dB reference standard and outperforms comparable designs in the literature. The proposed filter finds direct application in modern multi-band wireless systems including 4G LTE, UMTS, and sub-3 GHz 5G networks.

The most immediate extension is the physical fabrication and measurement of the prototype, allowing measured responses to be compared against simulation while accounting for substrate-thickness variation, metal surface roughness, and SMA connector effects. The transformation technique can also be extended to quad-band and higher-order designs, and the resonators can be miniaturised further using the folded-arms square open-loop resonator (FASOLR) reported in [3]. The filter can additionally be combined with filtering power-divider technology to form an integrated triple-band filtering power divider, and the operating frequencies can be reconfigured to target other bands such as 3.5 GHz 5G NR, 2.4 GHz Wi-Fi, and 1.8 GHz LTE by scaling the resonator dimensions.